\documentstyle[aps, array, manuscript]{revtex}
\tightenlines
\draft
\begin{document}
\title{Gauge-invariant metric fluctuations
from NKK theory of gravity: de Sitter expansion}
\author{
$^1$Jos\'e Edgar Madriz Aguilar\footnote{
E-mail address: edgar@itzel.ifm.umich.mx}
$^{2,3}$ Mariano Anabitarte\footnote{
E-mail address: anabitar@mdp.edu.ar}
and $^{2,3}$Mauricio Bellini\footnote{
E-mail address: mbellini@mdp.edu.ar}}
\address{
$^1$Instituto de F\'{\i}sica y Matem\'aticas,
AP: 2-82, (58040) Universidad Michoacana de San Nicol\'as de Hidalgo,
Morelia, Michoac\'an, M\'exico.\\
$^2$Departamento de F\'{\i}sica,
Facultad de Ciencias Exactas y Naturales,
Universidad Nacional de Mar del Plata,
Funes 3350, (7600) Mar del Plata, Argentina.\\
$^3$ Consejo Nacional de Ciencia y Tecnolog\'{\i}a (CONICET).}

\vskip .5cm
\maketitle
\begin{abstract}
In this paper we study gauge-invariant metric fluctuations
from a Noncompact Kaluza-Klein (NKK) theory of gravity in a
de Sitter expansion. We recover the well known result
$\delta\rho/\rho \simeq 2\Phi$, obtained from the standard
4D semiclassical approach to inflation. The spectrum for these
fluctuations should be dependent of the fifth (spatial-like) coordinate.
\end{abstract}
\vskip .2cm
\noindent
Pacs numbers: 04.20.Jb, 11.10.kk, 98.80.Cq \\
\vskip 1cm
\section{Introduction}

The relativistic theory of cosmological perturbations is a cornerstone in our
understanding of the early universe as it is indispensable in relating
early universe scenarios, such as inflation, to cosmological data
such as the Cosmic Microwave Background (CMB) anisotropies.
The inflationary model\cite{1'}
solves several difficulties which arise from the standard cosmological model,
such as the horizon, flatness and monopole problems, and it
provides a mechanism for the creation of primordial density of fluctuations,
nedeed to explain the structure formation\cite{starobinsky}.
The most
widely accepted 4D
approach assumes that the inflationary phase is driving by a
quantum scalar field $\varphi $ related to a scalar potential
$V(\varphi)$\cite{linde90}.
Within this
perspective, the semiclassical approach to
inflation proposes to describe the dynamics of
this quantum field on the basis of two pieces: the spatially
homogeneous (background)
and inhomogeneous components\cite{PRD96}.
Usually the homogeneous one is interpreted
as a classical field $\varphi_b(t) $
that arises from the vacuum expectation value of
the quantum field. The inhomogeneous component $\delta\varphi(\vec R,t)$
are the quantum fluctuations.
These quantum field fluctuations are responsible for metric fluctuations
around the background Friedmann-Robertson-Walker (FRW)
metric\cite{riotto}. 

The two current versions of 5D gravity theory are membrane theory\cite{1-3}
and induced-matter theory\cite{im}. In the former, gravity propagates
freely into the bulk, while the interactions of particle physics are confined
to a hypersurface (the brane). The induced-matter theory in its simplest
form is the basic Kaluza-Klein (KK) theory in which the fifth
dimension is not compactified and the field equations of general relativity
in 4D follow from the fact that the 5D manifold is Ricci-flat.
Thus the large
extra dimension is responsible for the appearance of sources in 4D
general relativity. Hence, the 4D world of general relativity is embedded
in a 5D Ricci-flat manifold.
There has recently been an uprising interest in finding exact solutions
of the Kaluza-Klein field equations in 5D, where the
fifth coordinate is considered as noncompact. This theory reproduces
and extends known solutions of the Einstein field equations in 4D. Particular
interest revolves around solutions which are not only
Ricci flat, but also Riemann flat. This is because it is
possible to have a flat 5D manifold which contains a curved
4D submanifold, as implied by the Campbell theorem. So, the
universe may be ``empty'' and simple in 5D, but contain matter of
complicated forms in 4D\cite{we}. 
This paper is devoted to study a 4D de Sitter expansion of the universe
from a NKK theory of gravity, taking into account the scalar metric
fluctuations, which are gauge-invariant.
In particular, we are aimed to describe the 4D dynamics of the
gauge-invariant scalar metric fluctuations from the NKK theory of
gravity. To make it we shall consider the action
\begin{equation} \label{action}
I=-\int d^{4}x \  d\psi\,\sqrt{\left|\frac{^{(5)}
\bar g}{^{(5)}\bar g_0}\right|} \ \left[
\frac{^{(5)}\bar R}{16\pi G}+ ^{(5)}{\cal 
L}(\varphi,\varphi_{,A})\right],
\end{equation}
for a scalar field $\varphi$, which is minimally coupled to gravity.
Since we are aimed to describe a manifold in apparent vacuum
the Lagrangian density ${\cal L}$ in (\ref{action}) should be only
kinetic in origin
\begin{equation}
^{(5)} {\cal L}(\varphi,\varphi_{,A}) = \frac{1}{2} g^{AB}
\varphi_{,A} \varphi_{,B},
\end{equation}
where $A,B$ can take the values $0,1,2,3,4$ and
the perturbed line element $dS^2=g_{AB} dx^A dx^B $ is given by
\begin{equation}\label{a}
dS^2 = \psi^2 \left(1+ 2 \Phi\right) dN^2 - \psi^2
\left(1- 2 \Psi\right) e^{2N} dr^2 - \left(1-Q\right) d\psi^2.
\end{equation}
Here, the fields $\Phi$, $\Psi$ and $Q$ are functions of the coordinates
[$N,\vec r(x,y,z),\psi$], where $N$, $x$, $y$, $z$ are dimensionless
coordinates and $\psi$ has spatial units. Note that
$^{(5)}\bar R$ in the action (\ref{action})
is the Ricci scalar evaluated on the background metric
$\left(dS^2\right)_b = \bar g_{AB} dx^A dx^B$. In our case we
shall consider the background canonical metric
\begin{equation}\label{back}
\left(dS^2\right)_b
= \psi^2 dN^2 - \psi^2 e^{2N} dr^2
-d\psi^2,
\end{equation}
which is 3D spatially isotropic, homogeneous and flat\cite{otro}.
Furthermore, the metric (\ref{back}) is globally flat (i.e.,
$\bar R^A_{BCD} =0$) and describes an apparent
vacuum: $\bar G_{AB}=0$.

The energy-momentum tensor is given by
\begin{equation}
T_{AB} = \varphi_{,A} \varphi_{,B}-
\frac{1}{2} g_{AB} \varphi_{,C} \varphi^{,C}.
\end{equation}

\section{Formalism}

In order to describe scalar metric fluctuations we
must consider the covariant energy-momentum tensor
$T_{AB}$ to be symmetric. In such a case
we obtain that $\Psi = \Phi$ and $Q=2\Phi$,
so that the line element (\ref{a}), now holds
\begin{equation}   \label{1}
dS^2
= \psi^2 \left(1+ 2 \Phi\right) dN^2 - \psi^2
\left(1- 2 \Phi\right) e^{2N} dr^2 - \left(1- 2 \Phi\right) d\psi^2,
\end{equation}
where the field $\Phi(N, \vec r, \psi)$ is the scalar metric perturbation
of the background 5D metric (\ref{back}).
For the metric (\ref{1}), $|^{(5)}\bar g|=\psi^8 e^{6N} $
is the absolute value of the determinant for 
the background metric (\ref{back})
and
$|^{(5)} \bar g_0|=
\psi^8_0 e^{6N_0}$
is a constant of dimensionalization, for the constants $\psi_0$ and $N_0$.
Furthermore,
$G=M^{-2}_p$ is the gravitational constant
and $M_p=1.2 \  10^{19} \  {\rm GeV}$ is the Planckian mass.
In this work we shall consider $N_0=0$, so that
$\left|^{(5)}\bar g_0\right|=\psi^8_0$.
Here, the index $`` 0 "$ denotes the value
at the end of inflation.

On the other hand, the contravariant metric tensor, after
a $\Phi$-first order approximation, is
\begin{equation}     \label{3}
g^{AB} = \left(\begin{array}{ccccc}
\frac{(1-2\Phi)}{\psi^2} & 0 & 0 & 0 & 0 \\
0 & -\frac{e^{-2N} (1+2\Phi)}{\psi^2} & 0 & 0 & 0 \\
0 & 0 & -\frac{ e^{-2N} (1+2\Phi)}{\psi^2} & 0 & 0 \\
0 & 0 & 0 & -\frac{ e^{-2N} (1+2\Phi)}{\psi^2} & 0 \\
0 & 0 & 0 & 0 & - (1+2\Phi)
\end{array}
\right),
\end{equation}
which can be written as $g^{AB} = \bar g^{AB} + \delta g^{AB}$, being
$\bar g^{AB}$ the contravariant background metric tensor.
The dynamics for $\varphi$ and $\Phi$ are well described
by the Lagrange and Einstein equations, which we shall study in the
following subsections.

\subsection{Lagrange equations}

The Lagrange equations for the fields $\varphi$ and $\Phi$ are
respectively given by
\begin{eqnarray}
&& \frac{\partial^2\varphi}{\partial N^2}
+ 3\frac{\partial\varphi}{\partial N} - e^{-2N} \nabla^2_r \varphi -
\psi\left(\psi \frac{\partial^2\varphi}{\partial\psi^2}
+ 4 \frac{\partial\varphi}{\partial \psi}\right)
\nonumber \\
&-&
2\Phi \left[\frac{\partial^2\varphi}{\partial N^2}
+ 3\frac{\partial\varphi}{\partial N} - e^{-2N} \nabla^2_r \varphi
+ \psi\left(\psi \frac{\partial^2\varphi}{\partial\psi^2}
+ 4 \frac{\partial\varphi}{\partial\psi}\right)\right]
- 2\left(\frac{\partial\varphi}{\partial N}
\frac{\partial\Phi}{\partial N} + \psi^2 \frac{\partial\Phi}{\partial\psi}
\frac{\partial\varphi}{\partial\psi} \right) =0,
\label{l1} \\
&& \left(\frac{\partial\varphi}{\partial N}\right)^2
+ e^{-2N} \left(\nabla\varphi\right)^2
+\psi^2 \left(\frac{\partial\varphi}{\partial\psi}\right)^2=0. \label{l2}
\end{eqnarray}
Now we can make the following semiclassical approximation:
$\varphi(N,\vec r, \psi) = \varphi_b(N,\psi) +
\delta\varphi(N,\vec r, \psi)$, such that $\varphi_b$
is the solution of eq. (\ref{l1})
in absense of the
inflaton and metric fluctuations [i.e.,
for $\Phi =
\nabla_r \varphi_b =0$].
Hence, the Lagrange equations for $\varphi_b$ and
$\delta\varphi$ are
\begin{eqnarray}
&& \frac{\partial^2\varphi_b}{\partial N^2}
+ 3 \frac{\partial\varphi_b}{\partial N}
-\psi \left[ \psi \frac{\partial^2\varphi_b}{\partial\psi^2} +
4 \frac{\partial\varphi_b}{\partial\psi} \right]=0, \label{l3} \\
&& \frac{\partial^2\delta\varphi}{\partial N^2}
+ 3 \frac{\partial\delta\varphi}{\partial N}
- e^{-2N} \nabla^2_r \delta\varphi - \psi \left[
4\frac{\partial\delta\varphi}{\partial\psi}
+ \psi \frac{\partial^2\delta\varphi}{\partial\psi^2} \right] \nonumber \\
& - & 2 \frac{\partial\varphi_b}{\partial N} \frac{\partial\Phi}{\partial N}
-2 \psi^2 \left[ \frac{\partial\varphi_b}{\partial\psi}
\frac{\partial\Phi}{\partial\psi}
+\left( \frac{\partial^2\varphi_b}{\partial\psi^2}
+\frac{4}{\psi} \frac{\partial\varphi_b}{\partial\psi}\right) \Phi\right] =0. \label{l4}
\end{eqnarray}
Note that for $\Phi= \nabla_r\varphi_b=0$, the
equation (\ref{l2}) holds
\begin{equation}\label{l5}
\left(\frac{\partial\varphi_b}{\partial N}\right)^2 + \psi^2
\left(\frac{\partial\varphi_b}{\partial\psi}\right)^2 =0,
\end{equation}
which will be useful later.

\subsection{5D Einstein equations}

The diagonal
perturbed first order 5D Einstein
equations $\delta G_{AA} = -8\pi G
\delta T_{AA}$, are
\begin{eqnarray}
&& 9 \frac{\partial \Phi}{\partial N} - 9\psi \frac{\partial
\Phi}{\partial \psi}
-3\psi^2 \frac{\partial^2 \Phi}{\partial\psi^2} -
3 e^{-2N} \nabla^2_r \Phi + 12 \Phi =
-16\pi G \psi^2 \Phi\left(\frac{\partial \varphi_b}{\partial\psi} \right)^2,
\label{e1} \\
&& 3\psi^2 e^{2N} \frac{\partial^2 \Phi}{\partial\psi^2} - 36 e^{2N} \Phi +
2\nabla^2_r \Phi - 30 e^{2N} \frac{\partial \Phi}{\partial N} - 9 e^{2N}
\frac{\partial^2 \Phi}{\partial N^2} + 3\psi \frac{\partial
\Phi}{\partial\psi}
= 48 \pi G e^{2N} \Phi \left(\frac{\partial\varphi_b}{\partial N}\right)^2,
\label{e2} \\
&& 3\frac{\partial^2 \Phi}{\partial N^2} - e^{-2N} \nabla^2_r \Phi +
24 \Phi + 15 \frac{\partial \Phi}{\partial N}
- 6\psi \frac{\partial \Phi}{\partial\psi} =
-16 \pi G \Phi \left(\frac{\partial\varphi_b}{\partial N}\right)^2,
\label{e3}
\end{eqnarray}
for the components $NN$, $rr$ and $\psi\psi$, respectively.
Furthermore, the non-diagonal 5D Einstein equations
(which are symmetric with respect to indices permutation):
$\delta G_{AB} = -8\pi G
\delta T_{AB}$ (with $A\neq B$), are
\begin{eqnarray}
&& \frac{\partial \Phi}{\partial x^i}
+ 3\frac{\partial^2 \Phi}{\partial x^i
\partial N}=0, \label{e4} \\
&& \psi \frac{\partial^2 \Phi}{\partial\psi\partial N} +
2 \frac{\partial \Phi}{\partial\psi} - \frac{\partial \Phi}{\partial N}=0,
\label{e5} \\
&& 3\frac{\partial \Phi}{\partial x^i} - \psi \frac{\partial^2
\Phi}{\partial x^i
\partial\psi} =0,
\end{eqnarray}
for the components $Nx^i$, $N\psi$ and $x^i\psi$, respectively
(latin indices can take values $1,2,3$).
After some algebra, from the equations (\ref{e1}), (\ref{e2}) and
(\ref{e3}), we obtain
\begin{equation}\label{eso}
\frac{\partial^2 \Phi}{\partial N^2}+ 3 \frac{\partial\Phi}{\partial N}
- e^{-2N} \nabla^2_r \Phi - 2 \psi^2 \frac{\partial^2\Phi}{\partial\psi^2}
+ \frac{16 \pi G}{3} \Phi \left[
\left(\frac{\partial\varphi_b}{\partial N}\right)^2 +
\psi^2 \left(\frac{\partial\varphi_b}{\partial\psi}\right)^2 \right] =0.
\end{equation}
From eq. (\ref{l5}), the eq. (\ref{eso}) holds
\begin{equation} \label{phi}
\frac{\partial^2 \Phi}{\partial N^2}+ 3 \frac{\partial\Phi}{\partial N}
- e^{-2N} \nabla^2_r \Phi - 2 \psi^2 \frac{\partial^2\Phi}{\partial\psi^2}
=0,
\end{equation}
which is the equation of motion for the 5D scalar metric fluctuations
$\Phi(N,\vec r, \psi)$.

\subsection{Normalization of $\Phi$ in 5D}

We consider the following separation of the 5D metric fluctuations:
$\Phi(N, \vec r, \psi) = \Phi_1(N) \Phi_2(\vec r) \Phi_3(\psi)$. 
The equation (\ref{phi}) can be rewritten as three differential equations
\begin{eqnarray}
&& \psi^2 \frac{d^2\Phi_3}{d\psi^2} = k^2_{\psi}\psi^2 \Phi_3,
\label{phi1} \\
&& \nabla^2_r \Phi_2 = - k_r^2 \Phi_2, \label{phi2} \\
&& \frac{d^2 \Phi_1}{dN^2} + 3 \frac{d\Phi_1}{dN} -
\left(2 k^2_{\psi}\psi^2 + e^{-2N} k_r^2 \right)  \Phi_1 =0, \label{phi3}
\end{eqnarray}
where $k^2_{\psi}\psi^2 > 0$.

Now we consider the transformation $\Phi(N,\vec r, \psi) =
e^{-3N/2} \left({\psi\over \psi_0}\right) \chi(N,\vec r)$.
This transformation also can be physically justified in the sense
that we can make observations on any hypersurface with constant
$\psi$.
The equation of motion for $\chi$ is
\begin{equation}\label{chi}
\frac{\partial^2 \chi}{\partial N^2} - \left(e^{-2N} \nabla^2_r
+2 k^2_{\psi}\psi^2\right)
\chi =0,
\end{equation}
where $\chi$ can be written as a Fourier expansion
\begin{equation}\label{fourier}
\chi(N,\vec r) = \frac{1}{(2\pi)^{3/2}} {\Large\int} d^3k_r {\Large\int}
dk_{\psi} \left[ a_{k_rk_{\psi}} e^{i\vec{k_r}.\vec r} \xi_{k_r k_{\psi}}(N)
+ a^{\dagger}_{k_rk_{\psi}} e^{-i\vec{k_r}.\vec r}
\xi^*_{k_r k_{\psi}}(N)\right],
\end{equation}
and the asterisk denotes the complex conjugate
and ($a_{k_Rk_{\psi}}$, $a^{\dagger}_{k_Rk_{\psi}}$) are, respectively,
the annhilation and creation operators which satisfy the algebra
\begin{displaymath}
\left[a_{k_rk_{\psi}}, a^{\dagger}_{k'_{r}k'_{\psi}}\right] =
\delta^{(3)}\left(\vec k_r - \vec k'_r \right) \delta\left(
\vec k_{\psi} - \vec k'_{\psi}\right), \quad
\left[a_{k_rk_{\psi}}, a_{k'_{r}k'_{\psi}}\right] =
\left[a^{\dagger}_{k_rk_{\psi}}, a^{\dagger}_{k'_{r}k'_{\psi}}\right] =0.
\end{displaymath}

The equation of motion for the $N$-dependent modes $\xi_{k_r k_{\psi}}$ is
\begin{equation}\label{xi}
\frac{d^2\xi_{k_r k_{\psi}}}{dN^2} + \left[ e^{-2N} k^2_r -
2k^2_{\psi} \psi^2\right] \xi_{k_r k_{\psi}}=0.
\end{equation}
The general solution for this equation is
\begin{equation}
\xi_{k_rk_{\psi}}(N) = C_1 \  {\cal H}^{(1)}_{\nu}[x(N)] +
C_2 \  {\cal H}^{(2)}_{\nu}[x(N)],
\end{equation}
where $\nu = \sqrt{2} k_{\psi} \psi$ is a constant
and $x(N) = k_r e^{-N}$.
Using the generalized Bunch-Davies vacuum\cite{bd}, we obtain that
\begin{equation}
\xi_{k_rk_{\psi}}(N) = i \sqrt{\frac{4}{\pi}} {\cal H}^{(2)}_{\nu} [x(N)],
\end{equation}
which are the normalized $N$-dependent modes of $\chi$.

\section{Effective 4D de Sitter expansion}

In this section we shall study the effective 4D $\Phi$-dynamics in an effective
4D de Sitter background expansion of the universe, which is considered
3D (spatially) flat, isotropic and homogeneous.

\subsection{Ponce de Leon metric}

We consider the transformation\cite{plb2005}
\begin{equation}\label{trans}
t = \psi_0 N, \qquad R=\psi_0 r, \qquad \psi = \psi.
\end{equation}
Hence, the 5D background metric (\ref{back}) becomes
\begin{equation}\label{pdl}
\left(dS^2\right)_b = \left(\frac{\psi}{\psi_0}\right)^2
\left[dt^2 - e^{2t/\psi_0} dR^2\right]-d\psi^2,
\end{equation}
which is the Ponce de Leon metric\cite{pdlm}, that describes a 3D spatially
flat, isotropic and homogeneous extended (to 5D) Friedmann-Robertson-Walker
metric in a de Sitter expansion.
Here, $t$ is the cosmic time and $R^2 = X^2+Y^2+Z^2$. 
This Ponce de Leon metric is a special case of the separable models studied
by him, and is an example of the much-studied class of canonical metrics
$dS^2 = \psi^2 g_{\mu\nu} dX^{\mu} dX^{\nu} - d\psi^2$\cite{MLW}.
Now we can take a foliation $\psi=\psi_0$ in the metric (\ref{pdl}), such
that the effective 4D metric results
\begin{equation}\label{desitter}
\left(dS^2\right)_b \rightarrow \left(ds^2\right)_b
= dt^2 - e^{2t/\psi_0} dR^2,
\end{equation}
which describes a 4D expansion
of a 3D spatially flat, isotropic and homogeneous universe that
expands with a constant Hubble parameter $H=1/\psi_0$ and a 4D
scalar curvature $^{(4)} {\cal R} = 12 H^2$.
Hence, the effective 4D metric of (\ref{1}) on hypersurfaces $\psi=1/H$, is
\begin{equation}\label{4d}
dS^2 \rightarrow ds^2 = \left(1+2\Phi\right) dt^2 -
\left(1-2\Phi\right) e^{2Ht} dR^2,
\end{equation}
where the metric (\ref{4d})
describes the perturbed 4D de Sitter expansion of the universe,
where $\Phi(\vec R,t)$ is gauge-invariant.

\subsection{Dynamics of $\Phi$ in an effective 4D de Sitter expansion}

In order to study the 4D dynamics of the
gauge-invariant scalar metric fluctuations $\Phi(\vec R,t)$
in a background de Sitter expansion we can take the equation
(\ref{phi}) with the transformations (\ref{trans}), for
$\psi=\psi_0=1/H$
\begin{equation}
\frac{\partial^2\Phi}{\partial t^2} + 3H \frac{\partial \Phi}{\partial t}
-e^{-2 H t} \nabla^2_R \Phi - \left.
2 \frac{\partial^2\Phi}{\partial\psi^2}\right|_{\psi=H^{-1}}=0,
\end{equation}
where
$\left.
\frac{\partial^2\Phi}{\partial\psi^2}\right|_{\psi=H^{-1}}=
k^2_{\psi_0} \Phi$.
To simplify the structure of this equation we propose
the redefined quantum metric fluctuations
$\chi(\vec R,t) = e^{3H t/2} \Phi(\vec R, t)$,
so that $\chi$ complies with the following equation of motion
\begin{equation}
\ddot\chi - e^{-2Ht} \nabla^2_R\chi - \left[
\frac{9}{4} H^2 + 4 k^2_{\psi_0}\right]\chi=0,
\end{equation}
where the redefined field $\chi(\vec R,t)$ can be expanded as
\begin{equation}
\chi(\vec R, t) = \frac{1}{(2\pi)^{3/2}} {\Large\int}d^3k_R {\Large\int}
dk_{\psi} \left[ a_{k_R k_{\psi}} e^{i \vec k_R.\vec R}
\xi_{k_R k_{\psi}}(t) + c.c.\right] \delta(k_{\psi}-k_{\psi_0}).
\end{equation}
Here, the operators $a_{k_Rk_{\psi}}$ and $a^{\dagger}_{k_R k_{\psi}}$
comply
\begin{displaymath}
\left[a_{k_Rk_{\psi}}, a^{\dagger}_{k'_{R}k'_{\psi}}\right] =
\delta^{(3)}\left(\vec k_R - \vec k'_R \right) \delta\left(
\vec k_{\psi} - \vec k'_{\psi}\right), \quad
\left[a_{k_Rk_{\psi}}, a_{k'_{R}k'_{\psi}}\right] =
\left[a^{\dagger}_{k_Rk_{\psi}}, a^{\dagger}_{k'_{R}k'_{\psi}}\right] =0.
\end{displaymath}
and the time dependent modes $\xi_{k_R k_{\psi_0}}(t)$ are given
by the equation
\begin{equation}
\ddot\xi_{k_R k_{\psi_0}}(t) + \left[k^2_R e^{-2Ht} - \left(\frac{9}{4} H^2+
4 k^2_{\psi_0}\right)\right]\xi_{k_R k_{\psi_0}}(t) =0.
\end{equation}
The general solution for this equation is
\begin{equation}
\xi_{k_R k_{\psi_0}}(t) = A_1 {\cal H}^{(1)}_{\mu}[y(t)] +
A_2 {\cal H}^{(2)}_{\mu}[y(t)],
\end{equation}
where $\mu = {\sqrt{9+16 k^2_{\psi_0}/H^2}\over 2}$
and $y(t) = {k_R e^{-H t}\over H}$.
Using the Bunch-Davies vacuum\cite{bd}, we obtain
\begin{equation}
\xi_{k_R k_{\psi_0}}(t) = i \sqrt{\frac{\pi}{4H}} {\cal H}^{(2)}_{\mu}[y(t)],
\end{equation}
which are the normalized time dependent modes of $\chi(\vec R,t)$.

\subsection{Energy density fluctuations}

In order to obtain the energy density fluctuations on the
effective 4D FRW metric, we must calculate
\begin{equation}
\frac{\delta\rho}{\left<\rho\right>}
=
\left.\frac{\delta T^N_N}{\left<T^N_N\right>}
\right|_{t=\psi_0 N, R=\psi_0 r, \psi=1/H},
\end{equation}
where $\delta T_{NN}=-{1\over 2} \delta g_{NN} \varphi_{,L}
\varphi^{,L}$ is linearized and the brackets $<...>$
denote the expectation value on the 3D hypersurface $R(X,Y,Z)$.
Using the semiclassical expansion $\varphi(\vec R, t)=
\varphi_b (t) + \delta\varphi(\vec R,t)$,
after some algebra we obtain
\begin{equation}
\frac{\delta\rho}{\left<\rho\right>} \simeq
2\Phi \left\{1-
\frac{\left<\left(\delta\dot\varphi\right)^2
+e^{-2Ht} \left(\nabla_R \delta\varphi\right)^2 +
2V(\delta\varphi)\right>}{\left(\dot\varphi_b\right)^2
+ 4 H^2 \left( \varphi_b\right)^2}\right\} \simeq 2\Phi,
\end{equation}
where we have considered the following approximation:
\begin{equation}\label{des}
\frac{\left<\left(\delta\dot\varphi\right)^2
+e^{-2Ht} \left(\nabla_R \delta\varphi\right)^2 +
2V(\delta\varphi)\right>}{\left(\dot\varphi_b\right)^2
+ 4 H^2 \left(\varphi_b\right)^2}
\ll 1,
\end{equation}
being $V(\delta\varphi)=V(\varphi) - V\left(\varphi_b\right)$
\begin{displaymath}
V(\delta\varphi) = -\frac{1}{2}\left[
\left.
g^{\psi\psi} \left(\frac{\partial\varphi}{
\partial\psi}\right)^2\right|_{\psi=H^{-1}} -
\left. \bar g^{\psi\psi}
\left(\frac{\partial \left(\varphi_b\right)}{\partial
\psi}\right)^2\right|_{\psi=H^{-1}}\right],
\end{displaymath}
with
\begin{equation}
V(\varphi_b) = -\frac{1}{2} \bar g^{\psi\psi} \left.\left(\frac{\partial
\varphi_b}{\partial\psi}\right)^2\right|_{\psi=H^{-1}}=
2H^2 \left(\varphi_b\right)^2.
\end{equation}
The approximation (\ref{des}) is valid during inflation on super Hubble
scales (on the infrared sector), on which the inflaton field fluctuations
are very ``smooth''.
Finally, we can compute the amplitude for the 4D gauge-invariant
metric fluctuations for a de Sitter expansion on the
infrared sector ($k_R \ll e^{Ht} H$)
\begin{equation}
\left<\Phi^2\right> = \frac{e^{-3Ht}}{(2\pi)^3}
{\Large\int}^{\epsilon e^{Ht} H}_0
d^3k_R \  \xi_{k_R} \xi^*_{k_R},
\end{equation}
where $\epsilon \simeq 10^{-3}$ is a dimensionless constant. The squared
$\Phi$-fluctuations
has a power-spectrum ${\cal P}(k_R)$
\begin{equation}
{\cal P}(k_R) \sim k^{3-\sqrt{9+16k^2_{\psi_0}/H^2}}_R,
\end{equation}
which is nearly scale invariant for $k^2_{\psi_0}\psi^2_0
=k^2_{\psi_0}/H^2 \ll 1$.
In other words, the 3D power-spectrum of the gauge-invariant metric
fluctuacions depends on the wavenumber $k_{\psi_0}$
related to the fifth coordinate on the hypersurface $\psi=\psi_0\equiv
H^{-1}$.

It is well known from experimental data\cite{PDG} that
\begin{equation}\label{condition}
n_s=0.97 \pm 0.03,
\end{equation}
where $n_s=4-\sqrt{9+16 k^2_{\psi_0}/H^2} $
is the energy perturbation spectral index.
From the experimental condition
(\ref{condition}), we obtain
\begin{equation}        \label{45}
0 \le k_{\psi_0} < 0.15 \  H,
\end{equation}
which is the main result of this paper.

\section{Final Comments}

In this paper we have studied 4D gauge-invariant metric fluctuations
from a NKK theory of gravity. In particular we have examined these
fluctuations in an effective 4D de Sitter expansion for the universe
using a first-order expansion for the metric tensor. A very important
result of this formalism is the confirmation of the well known 4D result
$\delta\rho/\rho \simeq 2\Phi$\cite{riotto},
during inflation. Furthermore, the
spectrum of the energy fluctuations depends on the fifth coordinate.
More exactly, the result (\ref{45}) can be written as
$\left(k_{\psi_0}\psi_0\right)^2 < (0.15)^2$, being
$\left( k_{\psi_0}\psi_0\right)^2$ the degenerated eigenvalue of the
equation (\ref{phi1}):
$\left.\psi^2 {\partial^2\Phi\over \partial\psi^2}
\right|_{\psi=\psi_0} =\left.
k^2_{\psi} \psi^2 \Phi\right|_{\psi=\psi_0}$, with
$t=\psi_0 N$, $R=\psi_0 r$ on the hypersurface $\psi=\psi_0=1/H$.
Of course, this formalism could be extended to other inflationary
and cosmological models where the expansion of the
universe is governed by a single scalar field.

\end{document}